\documentstyle[12pt]{article}
\def\Re{{\cal R \mskip-4mu \lower.1ex \hbox{\it e}\,}}
\def\Im{{\cal I \mskip-5mu \lower.1ex \hbox{\it m}\,}}
\def\ie{{\it i.e.}}
\def\eg{{\it e.g.}}

\def\etal{{\it et al.}}
\def\ibid{{\it ibid}.}
\def\sub#1{_{\lower.25ex\hbox{$\scriptstyle#1$}}}
\def\sul#1{_{\kern-.1em#1}}
\def\sll#1{_{\kern-.2em#1}}
\def\sbl#1{_{\kern-.1em\lower.25ex\hbox{$\scriptstyle#1$}}}
\def\ssb#1{_{\lower.25ex\hbox{$\scriptscriptstyle#1$}}}
\def\sbb#1{_{\lower.4ex\hbox{$\scriptstyle#1$}}}

\def\gev{\,{\rm GeV}}

\def\to{\rightarrow}
\def\mh{\ifmmode m\sbl H \else $m\sbl H$\fi}
\def\mch{\ifmmode m_{H^\pm} \else $m_{H^\pm}$\fi}
\def\mt{\ifmmode m_t\else $m_t$\fi}
\def\mc{\ifmmode m_c\else $m_c$\fi}
\def\mz{\ifmmode M_Z\else $M_Z$\fi}
\def\mw{\ifmmode M_W\else $M_W$\fi}
\def\mws{\ifmmode M_W^2 \else $M_W^2$\fi}
\def\mhs{\ifmmode m_H^2 \else $m_H^2$\fi}
\def\mzs{\ifmmode M_Z^2 \else $M_Z^2$\fi}
\def\mts{\ifmmode m_t^2 \else $m_t^2$\fi}
\def\mcs{\ifmmode m_c^2 \else $m_c^2$\fi}
\def\mchs{\ifmmode m_{H^\pm}^2 \else $m_{H^\pm}^2$\fi}
\def\ztwo{\ifmmode Z_2\else $Z_2$\fi}
\def\zone{\ifmmode Z_1\else $Z_1$\fi}
\def\mtwo{\ifmmode M_2\else $M_2$\fi}
\def\mone{\ifmmode M_1\else $M_1$\fi}
\def\tb{\ifmmode \tan\beta \else $\tan\beta$\fi}
\def\xw{\ifmmode x\sub w\else $x\sub w$\fi}
\def\ch{\ifmmode H^\pm \else $H^\pm$\fi}
\def\lum{\ifmmode {\cal L}\else ${\cal L}$\fi}
\def\inpb{\ifmmode ~{\rm pb}^{-1}\else $~{\rm pb}^{-1}$\fi}
\def\infb{\ifmmode ~{\rm fb}^{-1}\else $~{\rm fb}^{-1}$\fi}
\def\epem{\ifmmode e^+e^-\else $e^+e^-$\fi}
\def\ppb{\ifmmode \bar pp\else $\bar pp$\fi}

\def\half{\textstyle{{1\over 2}}}

\newskip\zatskip \zatskip=0pt plus0pt minus0pt
\def\matth{\mathsurround=0pt}
\def\lsim{\mathrel{\mathpalette\atversim<}}
\def\gsim{\mathrel{\mathpalette\atversim>}}
\def\atversim#1#2{\lower0.7ex\vbox{\baselineskip\zatskip\lineskip\zatskip
  \lineskiplimit 0pt\ialign{$\matth#1\hfil##\hfil$\crcr#2\crcr\sim\crcr}}}

\renewcommand{\thefootnote}{\fnsymbol{footnote}}

\begin{document} \begin{titlepage}
\setcounter{page}{1}
\thispagestyle{empty}
\rightline{\vbox{\halign{&#\hfil\cr
&ANL-HEP-PR-92-28\cr
&MAD/PH/698\cr
&March 1992\cr}}}
\vspace{1in}
\begin{center}

{\Large\bf
Signals for Virtual Leptoquark Exchange at HERA}
\medskip

\normalsize M.A. DONCHESKI$^{a}$ 
and
J.L. HEWETT$^{b,}$\footnote{Research supported by an SSC Fellowship from the
Texas National Research Laboratory Commission}

\medskip
$^a$ Department of Physics\\
University of Wisconsin\\ Madison, WI 53706\\
\smallskip
\smallskip
$^b$ High Energy Physics Division\\Argonne National
Laboratory\\Argonne, IL 60439\\

\end{center}

\begin{abstract}

We study the effects of virtual leptoquarks on charged current and neutral
current processes at the $ep$ collider HERA.  We present the areas of
parameter space that can be excluded at HERA by searching for deviations from
Standard Model expectations.   The best results are obtained by examining the
ratio of neutral current to charged current cross sections,
$R=\sigma_{NC}/\sigma_{CC}$, where, with $200\inpb$ of integrated luminosity
for unpolarized $e^-$ and $e^+$ beams, HERA can search for leptoquarks with
masses up to $\sim 800\gev$, with leptoquark coupling strengths of order
$\alpha_{em}$.

\end{abstract}

\renewcommand{\thefootnote}{\arabic{footnote}} \end{titlepage}


Many theories which go beyond the Standard Model (SM) are inspired by the
symmetry between the quark and lepton generations and try to relate them at a
more fundamental level.  As a result, many of these models contain new
particles, called leptoquarks, which naturally couple to a lepton-quark pair.
Examples of such theories include\cite{bigref} models with quark-lepton
substructure, the strong-coupling version of the SM, horizontal symmetries,
grand unified theories based on the gauge groups SU(5), SO(10), as well as the
Pati-Salam SU(4), and superstring inspired $E_6$ models.  In all these
theories, leptoquarks carry both baryon and lepton number, and are triplets
under SU(3)$_C$; their other quantum numbers (\eg, spin, weak isospin, and
electric charge) vary between the different models.  These particles need not
be heavy; in fact, leptoquarks can have a mass $\lsim 100\gev$ and still
avoid\cite{fcnc} conflicts with rapid proton decay and dangerously large
flavor changing neutral currents.  This is particularly true in models where
each generation of fermions has its own leptoquark(s) which couples only
within that generation.  Models where the leptoquarks can possess flavor
non-diagonal couplings are much more constrained.

By its nature, the high-energy ep collider HERA is especially well-suited to
study leptoquarks.  Direct production\cite{brw,rest} can occur via an
$s$-channel resonance at enormous rates with distinctive peaks in the
$x$-distribution.  Indeed, HERA experimental searches are
expected\cite{harnew} to reach a discovery limit (with a $5 \sigma$ signal and
an integrated luminosity of $200\inpb$) of $\sim 250\gev$ when the leptoquark
coupling strength is of order $0.01\alpha_{em}$, and up to the kinematic limit
if the coupling is equal to $\alpha_{em}$.  If leptoquarks are too massive to
be produced directly at HERA, perhaps they can be detected through their
indirect effects via virtual exchange by searching for deviations from SM
expectations for certain processes.  This is similar in nature to the
detection\cite{pep} of the SM $Z$-boson at PEP/PETRA.  Several
authors\cite{brw,tony} have examined such effects on the neutral current
asymmetries that can be formed with polarized electron beams, and have found
that departures from the SM are small, even for leptoquarks of low mass
({\it e.g.}, $\sim 400\gev$) and large couplings.  Here, we systematically
investigate these possible indirect effects on charged current as well as
neutral current processes.  We find that the best results are obtained by
examining the ratio $R=\sigma_{NC}/\sigma_{CC}$, where discovery limits can
reach leptoquark masses of order $800\gev$ for electromagnetic coupling
strengths (with $200\inpb$ of integrated luminosity per $e^+, e^-$ beam).
Taking the ratio of neutral to charged current cross sections is also
advantageous because several systematic uncertainties, as well as those from a
lack of detailed knowledge of the parton distributions, will cancel.  Such a
ratio has historically played a great role\cite{tgr} in our knowledge of SM
interactions from traditional low-energy neutrino scattering.  As such, it is
only natural that such a ratio be used to probe new interactions in the
high-energy regime at HERA.

Present bounds from the LEP experiments\cite{lep} on a leptoquark mass are
$m > M_Z/2$; single leptoquark production from $Z$ decay could extend\cite{tom}
the LEP search reach to masses of $\sim 70\gev$.  It is also possible, using
the high statistics available at LEP, to study the virtual effects of
leptoquarks via flavor changing $Z$ decays\cite{meandrickandcompany}.  Any
such limits will, however, be more model dependent that those quoted above.
At hadron colliders, leptoquarks may be produced\cite{sandip} either singly
through their unknown $q\ell(LQ)$ couplings via $qg$ fusion, or in pairs by
$gg$ fusion and $q\bar q$ annihilation.  UA2 has placed\cite{uatwo} limits on
scalar leptoquark pair production by searching for the possible final states\
$e^+e^- +2$ jets\ ~and~\ $e\nu +2$ jets\ with the result that $m > 67\gev$ at
$95\%$ C.L.  assuming the branching fraction $B(LQ\to eq)=50\%$.  As one
varies this branching fraction from $10\%$ to $100\%$, the UA2 bounds range
from the LEP result of $m>M_Z/2$ up to $m>74\gev$.  Constraints in the
leptoquark mass-coupling plane may also be obtained by examining $t$-channel
leptoquark exchange in the process $e^+e^-\to q\bar q$.  Such effects are
hidden at LEP I due to the $Z$-boson resonance, but data from PEP/PETRA
allow\cite{epem} for leptoquark coupling strengths to be of order
electromagnetic strength, or larger, for the mass range of interest here.

For definiteness, we will concentrate on the leptoquark present in
superstring-inspired $E_6$ models\cite{esix}.  These leptoquarks, which we
will denote by $S$, are scalar, charge $-1/3$, baryon number $=+1/3$, lepton
number $=+1$, weak iso-singlets, and are the supersymmetric partner of the
exotic color triplet fermion which appears in the {\bf 27} representation of
$E_6$.  Their interactions are governed by the $E_6$ superpotential terms
\begin{equation}
\lambda_LLS^cQ + \lambda_RSu^ce^c + \lambda'\nu^cSd^c \,,
\end{equation}
where $L$ and $Q$ represent the left-handed lepton and quark doublets,
respectively, and the superscript $c$ denotes the charge conjugate states.
The Yukawa couplings, \ie, the $\lambda$'s, are {\it a priori} unknown and for
simplicity we set $\lambda_L=\lambda_R$ in our numerical calculations.  We
make the assumption that the right-handed neutrino is too heavy to be produced
at HERA and hence ignore any contributions from the $\lambda'$ term.  If
$\nu^c$ is light enough to be produced, the signature from its subsequent
decay would be distinguishable from that of the SM charged current events
considered here.  The total leptoquark decay width (assuming the right-handed
neutrino does not contribute) is
\begin{equation}
\Gamma_S = {m_S\over 16\pi} (2\lambda_L^2+\lambda_R^2)  \,.
\end{equation}
For calculational purposes, we parameterize the $\lambda$'s by
\begin{equation}
{\lambda_{L,R}^2\over 4\pi} = F_{L,R}~~\alpha \,,
\end{equation}
and take $F_{L}=F_{R}\le 1$.

First we consider the contributions of leptoquark exchange on neutral current
(NC) events.  These $e^\pm  q\to e^\pm  q$ and $e^\pm\bar q\to e^\pm\bar q$
events are generated by $t$-channel $\gamma$ and $Z$ exchange in the SM, and
by $s$- and $u$-channel leptoquark exchange.  The Feynman diagrams responsible
for these mechanisms are presented in Fig.~1a-b.  The differential cross
section for a left-handed polarized electron and an unpolarized proton is
\begin{eqnarray}
{d\sigma(e^-_Lp)\over dxdy} & = & {2 \pi \alpha^2 \over s x^2 y^2}
\left( \sum_{q = u,d} [ b_{LL}^2 (q) + b_{LR}^2 (q) (1 - y)^2 ] x q(x)
\right. \\
& + & \sum_{q = u,d} [ b_{LR}^2 (q) + b_{LL}^2 (q) (1 - y)^2 ] x
\bar{q}(x)  \nonumber \\
& + & F_L b_{LL} (u)\; sxy \left\{ {(\hat{s} - m_{S}^2)  \; x u(x) \over
[(\hat{s} - m_{S}^2)^2 + (m_{S} \Gamma_{S})^2]} + { (1 - y)^2
\; x \bar{u}(x)
\over [u - m_{S}^2]} \right\} \nonumber \\
& + & \left. {1 \over 4} F_L (F_L + F_R) \left\{ {t^2 \; x u(x) \over
[(\hat{s} - m_{S}^2)^2 + (m_{S} \Gamma_{S})^2]} + {u^2 y^2 \; x \bar{u}(x)
\over [u - m_{S}^2]^2} \right\} \right) \,, \nonumber
\end{eqnarray}
and for a left-handed polarized positron and an unpolarized proton,
\begin{eqnarray}
{d\sigma(e^+_Lp)\over dxdy} & = & {2 \pi \alpha^2 \over s x^2 y^2}
\left( \sum_{q = u,d} [ b_{RL}^2 (q) + b_{RR}^2 (q) (1 - y)^2 ] x q(x)
\right. \\
& + & \sum_{q = u,d} [ b_{RR}^2 (q) + b_{RL}^2 (q) (1 - y)^2 ] x
\bar{q}(x)  \nonumber \\
& + & F_R b_{RR} (u)\; sxy \left\{ {(1 - y)^2 \; x u(x) \over [u - m_{S}^2]}
+ {(\hat{s} - m_{S}^2)  \; x \bar{u}(x) \over
[(\hat{s} - m_{S}^2)^2 + (m_{S} \Gamma_{S})^2]} \right\} \nonumber \\
& + & \left. {1 \over 4} F_R (F_L + F_R) \left\{ {u^2 y^2 \; x u(x) \over
[u - m_{S}^2]^2} + {t^2 \; x \bar{u}(x) \over
[(\hat{s} - m_{S}^2)^2 + (m_{S} \Gamma_{S})^2]} \right\} \right) \,. \nonumber
\end{eqnarray}
The corresponding cross sections for right-handed polarized $e$'s are obtained
by
\begin{equation}
{d\sigma(e^\pm_Rp)\over dxdy} = {d\sigma(e^\pm_Lp)\over
dxdy}~~~~~~{\rm with}~~~~~~L\leftrightarrow R \,.
\end{equation}
Here $q(x)$ and $\bar q(x)$ represent the quark and anti-quark distribution
functions, $\hat s=xs$, $t=-Q^2$, $u=-\hat s+Q^2$, $Q^2=sxy$, and $x$ and $y$
are the usual scaling variables.  The functions $b_{ij}$ are given by
\begin{equation}
b_{ij}=-Q^q+{\sqrt{2}G_FM_Z^2C_i^eC_j^q\over\pi\alpha}
{t(t-M_Z^2)\over (t-M_Z^2)^2+(M_Z\Gamma_Z)^2} \,,
\end{equation}
with $Q^q$ being the electric charge of the incoming parton, $C_L^f=T^f_{3L}
-Q^f\sin^2\theta_w$, $C_R^f=-Q^f\sin^2\theta_w$, and $T^f_{3L}$ is the value
of the fermion's third component of weak isospin.  These expressions agree
with those of Ref.~\cite{brw} and also with the SM contributions of
Ref.~\cite{steve,pdg}.

The Feynman diagrams representing the SM and s-, u-channel leptoquark
contributions to charged current (CC) events, $e^\pm q\to \nu q$ and $e^\pm
\bar q\to \nu\bar q$,  are depicted in Fig. 2a-b.  The differential cross
section for a left-handed polarized electron is
\newpage
\begin{eqnarray}
{d\sigma(e^-_Lp)\over dxdy} & = & {2 \pi \alpha^2 \over s x^2 y^2}
\Bigg( b_{LL}^2 x u(x) + b_{LL}^2 (1 - y)^2  x \bar{d}(x) \\
& + & F_L b_{LL} s x y \left\{ {(\hat{s} - m_{S}^2) \; x u(x)
\over [(\hat{s} - m_{S}^2)^2 + (m_{S} \Gamma_{S})^2]}
+ {(1 - y)^2 \; x \bar{d}(x) \over [u - m_{S}^2]}  \right\} \nonumber \\
& + & {F_L^2 t^2 \; x u(x) \over 4 [ (\hat{s} - m_{S}^2)^2
+ (m_{S} \Gamma_{S})^2 ]} + {F_L^2 u^2 y^2 \; x \bar{d}(x) \over 4 [u
- m_{S}^2]^2} \Bigg) \,, \nonumber
\end{eqnarray}
and for a right-handed polarized positron,
\begin{eqnarray}
{d\sigma(e^+_Rp)\over dxdy} & = & {2 \pi \alpha^2 \over s x^2 y^2}
\Bigg( b_{LL}^2 (1 - y)^2 x d(x) + b_{LL}^2 x \bar{u}(x) \\
& + & F_L b_{LL} s x y \left\{ {(1 - y)^2 \; x d(x) \over [u - m_{S}^2]}
+ {(\hat{s} - m_{S}^2) \; x \bar{u}(x) \over
[(\hat{s} - m_{S}^2)^2 + (m_{S} \Gamma_{S})^2]} \right\} \nonumber \\
& + & {F_L^2 u^2 y^2 \; x d(x) \over 4 [u - m_{S}^2]^2}
+ {F_L^2 t^2 \; x \bar{u}(x) \over 4 [(\hat{s} - m_{S}^2)^2 + (m_{S}
\Gamma_{S})^2]} \Bigg) \,. \nonumber
\end{eqnarray}
Here, the SM contributions agree with that in Ref.~\cite{pdg}.  The leptoquark
exchange diagrams also contribute to $e^-_R$ and $e^+_L$ scattering, (note
that there are no SM contributions to these processes), with
\begin{equation}
{d\sigma(e^-_Rp)\over dxdy} = {2 \pi \alpha^2 \over s x^2 y^2}
{F_L F_R \over 4} \left( {t^2 \; x u(x) \over
[ (\hat{s} - m_{S}^2)^2 + (m_{S} \Gamma_{S})^2 ]}
+ {u^2 y^2 \; x \bar{d}(x) \over [u - m_{S}^2]^2} \right) \,,
\end{equation}
and
\medskip
\begin{equation}
{d\sigma(e^+_Lp)\over dxdy} = {2 \pi \alpha^2 \over s x^2 y^2}
{F_L F_R \over 4} \left( {u^2 y^2 \; x d(x) \over [u - m_{S}^2]^2}
+ {t^2 \; x \bar{u}(x) \over [ (\hat{s} - m_{S}^2)^2 + (m_{S}
\Gamma_{S})^2 ]} \right) \,.
\end{equation}
\medskip
For charged current processes, $b_{LL}$ is defined as
\begin{equation}
b_{LL}={G_FM_W^2\over\sqrt 2\alpha\pi} {t(t-M_W^2)\over
(t-M_W^2)^2+(M_W\Gamma_W)^2} \,.
\end{equation}

In order to ensure that the NC and CC events are cleanly separated and
identified we impose cuts on the scaling variables $x$ and $y$, as well as on
the  transverse momentum of the out-going lepton.  We have found that
restricting $x$ to lie in the range $0.1\le x\le 1.0$ satisfies\cite{ing} the
above condition, while yielding the greatest sensitivity to the indirect
leptoquark effects.  For a given value of $x$, we then restrict $y$ to lie in
the range
\begin{equation}
{\rm max}(0.1, y_{min})\le y\le {\rm min}(1.0, y_{max}) \,,
\end{equation}
where,
\begin{equation}
y_{max,min}=\half \left[ 1 \pm \sqrt{ 1 - { \mbox{4($p_T^{cut})^2$} \over
\mbox{$xs$}}} \;\; \right] \,.
\end{equation}
In order to fully distinguish the two types of events, we impose the realistic
cuts\cite{wes}
\begin{equation}
p_T(e)>5\gev
\end{equation}
for the out-going $e^\pm$ in the NC events, and
\begin{equation}
\not p_T(\nu)>20\gev
\end{equation}
on the missing transverse momentum in CC events.  In the circumstance that a
more sophisticated detector triggering system is developed, we also ran a test
case with a cut on the CC missing transverse momentum of $\not p_T(\nu) >
10\gev$, and found that our results did not significantly change.  We then
integrate the above differential cross sections over these $x$ and $y$ ranges
to obtain our results.

We perform a $\chi^2$ analysis in searching for virtual leptoquark effects and
comparing them to SM expectations for various asymmetries, total cross
sections, and the ratio $R=\sigma_{NC}/\sigma_{CC}$.  We assume that most of
the systematic uncertainties cancel for the asymmetries and in $R$, and only
include the statistical errors for these quantities, which are given by
\begin{equation}
\delta A= {1-A^2\over\sqrt{1-A}}{1\over\sqrt{2N_\alpha}} \,,
\end{equation}
for the asymmetries defined as $A=(\sigma_\alpha - \sigma_\beta)/
(\sigma_\alpha + \sigma_\beta)$ with $N_\alpha={\cal L}\sigma_\alpha$ being
the number of events for $\sigma_\alpha$ for a value of the integrated
luminosity ${\cal L}$, and
\begin{equation}
{\delta R\over R}={1\over\sqrt{N_{NC}}}\oplus{1\over\sqrt{N_{CC}}} \,,
\end{equation}
where the `$\oplus$' signifies that the errors are to be added in quadrature,
and $N_{NC,CC}$ is the number of neutral current and charged current events,
respectively.  In the case where we examine possible deviations in the total
cross sections, we also include the systematic errors associated with the
luminosity determination, giving
\begin{equation}
{\delta\sigma\over\sigma}={1\over\sqrt N}\oplus{\delta{\cal L}
\over{\cal L}} \,,
\end{equation}
where we take\cite{wes} the value $\delta {\cal L}/{\cal L} = 5\%$.  Finally,
we also use the sum of neutral current and charged current asymmetries in
order to increase the statistics in searching for deviations from the SM.  In
this case, we calculate $\chi^2$ for the neutral current ($\chi^2_{NC}$) and
for the charged current ($\chi^2_{CC}$) processes separately, and then add
\begin{equation}
\chi^2_{TOT} = \chi^2_{NC} + \chi^2_{CC} \,.
\end{equation}
We then determine 90\% and 95\% C.L. deviations based on the value of
$\chi^2_{TOT}$.

We present our results for the HMRS-B parton distributions of
Ref.~\cite{hmrs}.  We have checked that the limits obtained from the
asymmetries and the ratio of cross sections are insensitive to the choice of
distribution functions as expected, but the bounds from the total cross
sections can change by as much as $10\%$ as the distribution function
parameterizations are varied.  In our numerical calculations, we use the
values\cite{carter} $M_Z = 91.175\gev$, $\Gamma_Z = 2.487\gev$, $M_W =
80.14\gev$, $\Gamma_W = 2.15\gev$, $\alpha(M_Z) = 1/128$, and $x_w = 0.233$.

We examine the effects using both unpolarized and polarized $e^\pm$ beams.  In
the latter case, we define the polarized cross sections as
\begin{eqnarray}
\sigma^\pm(+P) &= \half (1+P)\sigma(e^\pm_Lp)+\half (1-P)\sigma(e^\pm_Rp) \,,
\nonumber\\
\sigma^\pm(-P) &= \half (1-P)\sigma(e^\pm_Lp)+\half (1+P)\sigma(e^\pm_Rp) \,,
\end{eqnarray}
which takes into account that there will be only a finite degree of
polarization.  We take $P=80\%$ and assume $125\inpb$ of integrated luminosity
for each polarized $e^\pm$ beam.  For the unpolarized case, we set $P=0$ in
the above equations.  Generally, we find that the use of polarized beams does
not increase the leptoquark search reach over the unpolarized case (for a
similar value of ${\cal L}$).

We begin the discussion of our results by examining the six different neutral
current asymmetries that can be formed from various combinations of the
polarized $e^\pm$ beams.  Out of these six asymmetries, $A^{-+}_{LR}$, which
is defined as
\begin{equation}
A^{-+}_{LR}={\sigma^-(+P)-\sigma^+(-P)\over \sigma^-(+P)+\sigma^+(-P)}\,,
\end{equation}
yields the best leptoquark search limits.  These limits are shown in the
leptoquark coupling-mass parameter plane in Fig. 3,  represented by the dotted
(dashed-dotted) curve in the upper left-hand corner which corresponds to the
$90\%~(95\%)$ C.L.  Here, and in all of our results below, the discovery
region lies to the left of the curves.  It is clear from the figure that these
types of measurements do {\bf not} yield useful limits, reaching leptoquark
masses of only $m_S\sim 325\gev$, even with electromagnetic coupling
strengths.  This confirms the results of previous authors\cite{brw,tony}.

We next discuss the limits that are obtainable from measurements of the total
cross sections.  Summing over both the neutral and charged current unpolarized
cross sections, we find that the $90\%~(95\%)$ C.L. discovery reach lies to
the left of the solid (dashed) curves in Fig. 3, for $200\inpb$ of integrated
luminosity per $e^-,e^+$ beam.  We find that the $5\%$ uncertainty in the
luminosity determination completely dominates the statistical errors, such
that improved searches from this technique are not possible, even with
substantial increases in the total integrated luminosity.  Of course,
measurements involving only cross sections are also severely hampered by the
rest of the systematic errors as well as the uncertainties arising from the
lack of knowledge of the parton distributions.

A slightly more useful quantity in searching for indirect leptoquark effects
is the (unpolarized) charge asymmetry $A^{-+}$, given by
\begin{equation}
A^{-+}={\sigma(e^-p)-\sigma(e^+p)\over\sigma(e^-p)+\sigma(e^+p)} \,.
\end{equation}
In Fig. 4 we show the $90\%$ (solid curves) and $95\%$ (dashed curves) C.L.
discovery regions obtainable from the charge asymmetry for (a) $200\inpb$ and
(b) $500\inpb$ of integrated luminosity per beam.  Where possible, we present
the results from neutral current and charged current events separately, as
well as those obtained from the sum of both types of events.  As can be seen
by comparing Figs. 3 and 4, the charge asymmetry can exclude a slightly larger
region of parameter space than cross section measurements alone.

It is, of course, possible to measure other quantities that are essentially
independent of systematic uncertainties.  We find two  such quantities that
provide much stronger search limits than any of the observables discussed so
far.  The first of these is the neutral current-charged current asymmetry,
defined as
\begin{equation}
A_{ZW} = \frac{\sigma_{NC} - \sigma_{CC}}{\sigma_{NC} + \sigma_{CC}}\,,
\end{equation}
and the second is the ratio $R=\sigma_{NC}/\sigma_{CC}$.  Here, $\sigma_{NC}$
and $\sigma_{CC}$ represent the total neutral current and charged current
cross sections, summed over both electron and positron beams.  The discovery
regions in the leptoquark coupling-mass plane, based on measurements of these
two observables, are displayed in Figs. 5 and 6 (note the scale change on the
$m_{S}$ axis compared to Figs. 3 and 4).  Figures 5 and 6a show the results
from unpolarized $e^\pm$ beams for integrated luminosities of $20,~200,$ and
$500\inpb$ per beam, while Fig. 6b presents the bounds obtainable from
polarized beams, with $125\inpb$ per $e^\pm_{L,R}$ beam.  Comparing Figs. 6a
and 6b, we see that although the total integrated luminosity (summed over all
beams) is larger in the polarized case, it is the unpolarized measurements
which yield the largest discovery region!  Note also that the search regions
resulting from measurements of $R$ and $A_{ZW}$ are essentially equivalent.
However, due to slightly different systematic uncertainties associated with
identifying NC versus CC events, the asymmetry $A_{ZW}$ may not be as clean a
measurement as the ratio $R$.  The region that can be explored with $200\inpb$
per unpolarized beam is $m_{S} \lsim 800\gev$ for large
leptoquark-electron-quark couplings ($F \sim  1$), and $F \gsim 0.13$ for
$m_{S} \sim 314\gev$, based on measurements of these two quantities.

In summary, we have performed a careful analysis of the effects of virtual
leptoquarks at HERA.  We found that measurements of the standard neutral
current asymmetries using polarized beams, of the charge asymmetries, or of
the neutral current and charge current total cross sections are either plagued
by systematic uncertainties or are simply not very sensitive to the parameters
of the theory.  However, it is possible, by examining ratios and asymmetries
between NC and CC cross sections, to eliminate the large systematic errors and
to probe a sizable region of the leptoquark parameter space.  In principle,
HERA can search for leptoquarks up to a mass of $800\gev$ for large
leptoquark-electron-quark couplings, and for leptoquark-electron-quark
couplings $\gsim 0.13 \alpha_{em}$ for leptoquarks with mass $m_S>\sqrt s$.

\vskip.25in
\centerline{ACKNOWLEDGEMENTS}

The research of J.L.H. was supported in part by awards granted by the Texas
National Research Laboratory Commission and by the U.S.~Department of Energy
under contract W-31-109-ENG-38.  The research of M.A.D. was supported in part
by the U.S.~Department of Energy under contract DE-AC02-76ER00881, in part by
the Texas National Research Laboratory Commission under grant no. RGFY9173 and
in part by the University of Wisconsin Research Committee with funds granted
by the Wisconsin Alumni Research Foundation.

\newpage

%
\def\MPL #1 #2 #3 {Mod.~Phys.~Lett.~{\bf#1},\ #2 (#3)}
\def\NPB #1 #2 #3 {Nucl.~Phys.~{\bf#1},\ #2 (#3)}
\def\PLB #1 #2 #3 {Phys.~Lett.~{\bf#1},\ #2 (#3)}
\def\PR #1 #2 #3 {Phys.~Rep.~{\bf#1},\ #2 (#3)}
\def\PRD #1 #2 #3 {Phys.~Rev.~{\bf#1},\ #2 (#3)}
\def\PRL #1 #2 #3 {Phys.~Rev.~Lett.~{\bf#1},\ #2 (#3)}
\def\RMP #1 #2 #3 {Rev.~Mod.~Phys.~{\bf#1},\ #2 (#3)}
\def\ZPC #1 #2 #3 {Z.~Phys.~{\bf#1},\ #2 (#3)}
\def\IJMP #1 #2 #3 {Int.~J.~Mod.~Phys.~{\bf#1},\ #2 (#3)}

\newpage
{\bf FIGURE CAPTIONS} \\
\begin{itemize}

\item[Figure 1.]{Feynman diagrams contributing to the neutral current events
(a) $e^\pm q\to e^\pm q$ and (b) $e^\pm\bar q\to e^\pm\bar q$, in the Standard
Model and with leptoquark exchange.}

\item[Figure 2.]{Feynman diagrams contributing to the charged current events
(a) $e^-q\to \nu q'$ and (b) $e^+q\to\nu q'$ in the Standard Model and with
leptoquark exchange.}

\item[Figure 3.]{Discovery region in the $m_{S}-F$ plane at HERA by neutral
current plus charged current total cross section and by the asymmetry
$A_{LR}^{+-}$, for the integrated luminosities as shown.  The 90\% (95\%) C.L.
discovery region corresponds to the area to the left of the solid (dashed)
curves for the cross section measurements and to the dotted (dashed-dotted)
for the asymmetry.}

\item[Figure 4.]{Discovery region in the $m_{S}-F$ plane at HERA by the charge
asymmetry, $A^{-+}$, for the charged current ($A^{-+}_{CC}$), the neutral
current ($A^{-+}_{NC}$) and the sum ($A^{-+}_{CC} + A^{-+}_{NC}$) for (a)
${\cal L}=200\inpb$ and (b) $500\inpb$.  The area to the upper left of the
solid (dashed) curves can be excluded at the 90\% (95\%) C.L. }

\item[Figure 5.]{Search region in the $m_{S}-F$ plane at HERA by the
neutral-charged current asymmetry, $A_{ZW}$ for various values of ${\cal L}$
as shown.  The area to the upper left of the solid (dashed) curves can be
excluded at the 90\% (95\%) C.L.}

\item[Figure 6.]{Discovery region in the $m_{S}-F$ plane at HERA from the
ratio, $R$, with (a) unpolarized $e^\pm$ and (b) polarized $e^\pm$ beams, for
different integrated luminosities as shown.  The area to the upper left of the
solid (dashed) curves can be excluded at the 90\% (95\%) C.L.}

\end{itemize}

\begin{thebibliography}{99}
\bibitem{bigref}
S. Pakvasa, \IJMP A2 1317 1987 ; B. Schremp and F. Schremp, \PLB 153B 101
1985 ;
L.F. Abott and E. Farhi, \PLB 101B 69 1981 ;  \NPB B189 547 1981 ;  H. Georgi
and S.L. Glashow, \PRL 32 438 1974 ;  G. Senjanovi\`c and A. Sokorac, \ZPC C20
255 1983 ;  J.C. Pati and A. Salam, \PRD D10 275 1974 ;  E. Witten, \NPB B258
75 1985 ;  M. Dine \etal, \ibid, {\bf B259}, 519 (1985);  J. Breit, B.A.
Ovrut, and G. Segre, \PLB 158B 33 1985 .
\bibitem{fcnc}
See, for example, W. Buchm\"uller and D. Wyler, \PLB B177 377 1986 .
\bibitem{brw}
W. Buchm\"uller, R. R\"uckl, and D. Wyler, \PLB B191 442 1987 .
\bibitem{rest}
V. Angelopoulos \etal, \NPB B292 59 1987 ;  J.F. Gunion and E. Ma \PLB B195
257 1987 .
\bibitem{harnew}
N. Harnew, in {\it Proceedings of the 1987 DESY Workshop on HERA Physics},
Hamburg, Germany, October 1987.
\bibitem{pep}
R. Hollebeek, in Proceedings of the {\it 1981 International Symposium on
Lepton and Photon Interactions at High Energy}, Bonn, West Germany, August
1981, edited by W. Pfeil (Physikalisches Institut, Bonn, 1981);  Ch. Berger,
{\it et al.}, \ZPC C7 289 1981 ;  W. Bartel, {\it et al.}, \PLB 108B 140 1982 .
\bibitem{tony}
J.A. Grifols and S. Peris, \PLB B201 287 1988 ;  M.A. Doncheski and J.L.
Hewett, in Proceedings of the {\it 1990 Summer Study on High Energy Physics -
Research Directions for the Decade}, Snowmass, CO, 1990, ed. by E. Berger
(World Scientific, Singapore, 1992).
\bibitem{tgr}
For a historical perspective, see, for example, K. Winter, in {\it Neutrino
Physics}, edited by K. Winter (Cambridge University Press, London, 1991) and
references therein;  T.G. Rizzo, Ph.D. thesis, University of Rochester,
Rochester, NY (1978).  For a more recent summary, see U. Amaldi \etal, \PRD
D36 1385 1987 .
\bibitem{lep}
D. Decamp \etal, ALEPH Collaboration, CERN Report, CERN-PPE/91-149 (1991),
(submitted to {\it Phys. Rep.});  P. Abreu \etal, DELPHI Collaboration,
CERN Report, CERN-PPE-91-138 (1991);  B. Adeva \etal, L3 Collaboration, \PLB
B261 169 1991 ;  G. Alexander \etal, OPAL Collaboration, \PLB B263 123 1991 .
\bibitem{tom}
T.G. Rizzo, \PRD D44 186 1991 .
\bibitem{meandrickandcompany}
M.A. Doncheski, {\it et al.}, \PRD D40 2301 1989 .
\bibitem{sandip}
J.L. Hewett and S. Pakvasa \PRD  D37 3165 1988 ;  V. Barger, K. Hagiwara, T.
Han, and D. Zeppenfeld, \PLB B220 464 1989 .
\bibitem{uatwo}
J. Alitti, \etal, UA2 Collaboration, CERN Report, CERN-PPE/91-158.
\bibitem{epem}
J.L. Hewett and T.G. Rizzo, \PRD D36 3367 1987 ;  H. Dreiner \etal, \MPL A3
443 1988 .
\bibitem{esix}
For a review, see, J.L. Hewett and T.G. Rizzo, \PR 183 193 1989 .
\bibitem{steve}
S. Capstick and S. Godfrey, \PRD D35 3351 1987 .
\bibitem{pdg}
J. Hern\'andez \etal, Particle Data Group, \PLB B239 1 1990 .
\bibitem{ing}
G. Ingelman, D. Notz, and E. Ros, in {\it Proceedings of the 1987 DESY
Workshop on HERA Physics}, Hamburg, Germany, October 1987.
\bibitem{wes}
M. Derrick and W. Smith, private communication.
\bibitem{carter}
J. Carter talk presented at the {\it Joint International Lepton Photon
Symposium and Europhysics Conference on High Energy Physics}, Geneva,
Switzerland, July 1991 ;  H. Plothow-Besch, {\it ibid.} ;  J. Alitti \etal,
(UA2 Collaboration), \PLB B276 354 1992 , 365 (1992).
\bibitem{hmrs}
P.N. Harriman, A.D. Martin, W.J. Stirling and R.G. Roberts, \PRD D42 798 1990.
\end{thebibliography}
\end{document}